\documentstyle[preprint,aps]{revtex}

\begin{document}

\draft

\title{Semiclassical effects in black hole interiors}

\author{William A.\ Hiscock\cite{His} and Shane L.\ Larson\cite{Lar}}

\address{Department of Physics, Montana State University, Bozeman,
	Montana 59717}

\author{Paul R. Anderson\cite{And}}

\address{Department of Physics, Wake Forest University, Winston-Salem,
	NC 27109}

\date{January 2, 1997}

\maketitle
\begin{center}
MSUPHYS-97-001
\end{center}
\begin{abstract}

First-order semiclassical perturbations to the Schwarzschild black
hole geometry are studied within the black hole interior.  The source
of the perturbations is taken to be the vacuum stress-energy of
quantized scalar, spinor, and vector fields, evaluated using analytic
approximations developed by Page and others (for massless fields) and
the DeWitt-Schwinger approximation (for massive fields).  Viewing the
interior as an anisotropic collapsing cosmology, we find that
minimally or conformally coupled scalar fields, and spinor fields,
decrease the anisotropy as the singularity is approached, while vector
fields increase the anisotropy.  In addition, we find that massless
fields of all spins, and massive vector fields, strengthen the
singularity, while massive scalar and spinor fields tend to slow the
growth of curvature.

\end{abstract}

\pacs{ }

\section{INTRODUCTION}

The application of quantum field theory to curved space has resulted
in a large array of interesting and important results.  These include
black hole evaporation\cite{Hawk1} and its implications for black hole
thermodynamics\cite{GH} , the dissipation of anisotropy by particle
production in cosmological
spacetimes\cite{Z,ZS,H1,HFP,FPH,BB,LS,HP,HH}, and the removal of
cosmological singularities by vacuum polarization
effects\cite{PF,S,FHH,A,AZ}.  One of the places for which quantum
effects have been studied the least is the interior of a black hole.
One might think that such studies are not interesting because no
observer from the exterior region can probe the interior region unless
they choose to fall into the hole.  However the existence of black
hole evaporation makes it quite possible to eventually learn about
quantum effects in the interior of a black hole\footnote{By interior
we mean here the region inside the apparent horizon.}.  This is
because as a black hole evaporates more and more of its interior is
exposed.  Thus not only can quantum effects in the interior of a black
hole eventually be detected, they may have a significant influence on
the evaporation process.

Quantum effects in the interior may in fact have a direct bearing on
two of the most fundamental outstanding issues relating to the quantum
mechanics of black holes.  One of these is the question of what
happens during the late stages of black hole evaporation, that is,
what is the end point of the evaporation process?  The other is the
question of what happens to the information about how the black hole
formed.  There are at least two ways in which quantum effects in the
interior could affect the answers to these questions.  One is that if
quantum effects remove the singularity predicted by general relativity
then it is very likely that the evolution will be unitary and
information will not be destroyed.  A second possibility is that
quantum effects could cause the evaporation process to cease leaving a
zero temperature black hole remnant.  If the remnant has an event
horizon the information would very likely be trapped inside the black
hole.  Since the temperature of a black hole is determined by the
surface gravity at its horizon and since the evaporation process
causes the horizon to be at points which were previously in the
(apparent) interior, it is clear that the geometry of the interior is
likely to influence the evaporation process as it progresses.

One interesting quantum effect that seems likely to occur inside the
horizon of a black hole is the dissipation of anisotropy and possibly
inhomogeneity due to particle production.  This is because the
interior of such a black hole can be thought of as an anisotropic and
possibly inhomogeneous cosmology.  For example the interior of a
Schwarzschild black hole can be thought of as a homogeneous,
anisotropic cosmology of the Kantowski-Sachs family\cite{KS}.  It has
been well established that particle production dissipates anisotropy
in Bianchi Type I spacetimes\cite{Z,ZS,H1,HFP,FPH,BB,LS,HP,HH}.  If
the process of anisotropy dissipation occurs it will certainly alter
the geometry in the interior of a black hole.

For these reasons it is interesting to examine quantum effects in the
interior of a black hole.  To do so for either the interior or
exterior of an evaporating black hole would be an enormously difficult
task at present due to problems that one would encounter in computing
the stress-energy tensors for quantized fields in the relevant
spacetime.  However, computing the stress-energy tensors for these
fields in the case of a spherically symmetric black hole in thermal
equilibrium with radiation in a cavity, {\it i.e.,} with the fields in
the Hartle-Hawking state, is a much more tractable problem.  The
reason is that there are then three Killing vector fields in the
spacetime, which makes the mode equations separable.

For a black hole in equilibrium with fields in the Hartle-Hawking
state, analytical approximations for the stress-energy tensors of
various types of quantized fields have been obtained.  The derivations
of most of these approximations have been for the exterior region,
but, as is discussed later, they all can easily be extended to the
interior region.  These approximations include those of Page, Brown,
and Ottewill\cite{P,BO,PBO} for conformally invariant fields in
Schwarzschild spacetime, that of Frolov and Zel'nikov\cite{FZ1} for
conformally invariant fields in a general static spacetime, that of
Anderson, Hiscock and Samuel\cite{AHS} for massless arbitrarily
coupled scalar fields in a general static spherically symmetric
spacetime, and the DeWitt-Schwinger approximation for massive fields
which was derived by Frolov and Zel'nikov\cite{FZ82,FZ84} for Kerr
spacetime, by Anderson Hiscock and Samuel\cite{AHS} for a scalar field
in a general static spherically symmetric spacetime and most recently
by Herman and Hiscock\cite{RH} for an arbitrary spacetime.

In this paper the various approximations mentioned above are used to
investigate quantum effects in the interior of a Schwarzschild black
hole when the fields are in the Hartle-Hawking state.  The resulting
semiclassical backreaction equations are linearized about the
classical geometry and their solutions are found.  The questions of
whether backreaction effects tend to isotropize the spacetime and
whether they tend to ``soften'' the geometry as the singularity is
approached are addressed.  Although the questions of whether the
anisotropy is completely dissipated or whether the singularity is
removed cannot be answered by examining linear perturbations, the
results do provide insight into these issues.

In Section II the interior geometry of a Schwarzschild black hole is
reviewed and in Section III the various analytical approximations are
reviewed and discussed.  Solutions to the linearized backreaction
equations which are derived using these approximations are displayed
in Section IV.  In Section V the dissipation of anisotropy is computed
and in Section VI the change in the curvature is computed.  The
results are summarized and discussed in Section VII.

\section{SCHWARZSCHILD BLACK HOLE INTERIOR}

The Schwarzschild black hole is described by the metric:
\begin{equation}
    ds^2 = - \left( 1 - {2M \over r}  \right) dt^2 + \left( 1 -
    {2M \over r}  \right)^{-1} dr^2 + r^2 d{\Omega}^2 ,
    \label{SchMetric}
\end{equation}
where $d{\Omega}^2$ is the metric of the two-sphere.  The coordinate
$r$ runs from 0 to $\infty$, and $t$ from $-\infty$ to $+\infty$.  We
are thus considering the complete Schwarzschild manifold, as is
appropriate with the Hartle-Hawking vacuum state.  The black hole
interior is the region in which $0 \leq r \leq 2M$.  In the interior,
the vector field $\partial/\partial r$ is timelike and the vector
field $\partial/\partial t$ is spacelike; hence, the coordinate $t$ is
a spatial coordinate, while $r$ is a time coordinate.

The nature of the interior is more easily visualized if new coordinate
names are adopted to reflect the physical nature of the coordinates in
the region of interest.  Defining new coordinates by setting
\begin{equation}
   T \equiv r \qquad, \qquad x \equiv t ,
    \label{CoordNames}
\end{equation}
the metric takes the form
\begin{equation}
   ds^2 = - \left( {2M \over T} - 1 \right)^{-1} dT^2 + \left( 
   {2M \over T} - 1 \right) dx^2 + T^2 d{\Omega}^2 .
   \label{InsideMetric}
\end{equation}
The metric given by Eq.  (\ref{InsideMetric}) is clearly an
anisotropic homogeneous cosmology.  The vector field
$\partial/\partial t$ is, in the interior, one of the spacelike
Killing vector fields (along with those on the two-sphere) which
guarantee spatial homogeneity.  The spatial coordinate $x$ here runs
from $-\infty$ to $+\infty$, while $T$ runs from $2M$ down to zero at
the curvature singularity in the black hole interior.

The Schwarzschild manifold contains both an anisotropic expanding
universe, the ``white hole'' portion of the extended geometry, and an
anisotropic collapsing universe, the black hole interior.  In this
paper we shall base our discussion on the black hole interior portion
of the geometry, but all conclusions may be restated in terms of the
expanding white hole geometry due to the time reversal symmetry of
both the Schwarzschild geometry and the Hartle-Hawking state we shall
use to perturb it.  However the boundary conditions for the fields in
the two cases are very different.  In the black hole case they are 
``initial'' conditions, while in the white hole case they are ``final'' 
conditions for the interior region.

While it is conventional to write homogeneous cosmological metrics
in terms of a proper time coordinate, i.e., 
\begin{equation}
   \tau = \int {dT \over {\left( {2M \over T} - 1 \right)^{1/2}}}
    \quad .
   \label{ProperTime}
\end{equation}
in the present case the spatial metric components cannot be expressed
in closed algebraic form in terms of such a coordinate.  Upon carrying
out the integral in Eq.  (\ref{ProperTime}), one finds that the range
of the coordinate $T$ from $2M$ down to $0$ corresponds to an interval
of proper time equal to $\pi M$.

The spacetime described by the metric of Eq.  (\ref{InsideMetric}),
viewed as a cosmological model, is an anisotropic but homogeneous
spacetime in which (as $T$ proceeds from $2M$ down to zero) two
spatial dimensions are collapsing while one is expanding.  The
interior Schwarzschild cosmology is a special case of a Type I
Kantowski-Sachs model\cite{KS}.

Since the Schwarzschild metric is a vacuum solution, there is no
naturally defined four-velocity of cosmological ``matter''; however,
to explore the properties of the solution as an anisotropic cosmology,
it is helpful to define a set of fiducial geodesic observers with
four-velocities given by
\begin{equation}
   u^{\alpha} = \left( \left( {2M \over T} - 1 \right)^{1/2},0,0,0 
   \right) .
   \label{FourVel}
\end{equation}
These observers travel along world lines with $x$, $\theta$, and
$\phi$ constant.  In terms of the conserved quantities normally used
to describe geodesics in the exterior Schwarzschild metric, these
observers have zero angular momentum and zero energy at infinity.

The proper volume of a cube defined by a set of fiducial observers at
the corners, separated by coordinate distances $\Delta x$, $\Delta
\theta$, and $\Delta \phi$ is given by
\begin{equation}
   V \left( T \right) = \left( {2M \over T} - 1 \right)^{1/2} T^2 
   \Delta x \Delta \theta \Delta \phi \qquad .
   \label{ProperVolume}
\end{equation}
Since the fiducial observers have four-velocities given by Eq.
(\ref{FourVel}), the quantities $\Delta x$, $\Delta \theta$, and
$\Delta \phi$ are constant.  The volume goes to zero at both $T = 0$
and $T = 2M$.

Near the singularity at $T = 0$, the Schwarzschild metric of Eq.
(\ref{InsideMetric}) may be put into a form which is locally
asymptotic to a Kasner universe.  Let coordinates $y$ and $z$ be
defined as functions of and locally in the neighborhood of a point
$\left( \theta_0, \phi_0 \right)$ by
\begin{equation}
   y = 2 M (\theta-\theta_0) \qquad , \qquad z = 2 M sin \left(
   \theta_0 \right) (\phi-\phi_0) \quad .
   \label{yzCoords}
\end{equation}
While these coordinates cannot be extended to cover the two-sphere
they are perfectly adequate to describe the expansion and contraction
of the cosmology in a local neighborhood.  Near the singularity, the
Schwarzschild metric then takes the form of a Kasner universe with
exponents $p_1 = -1/3$, $p_2 = p_3 = 2/3$:
\begin{equation}
    ds^2 = - d\tau^2 + \left( {\tau \over \tau_0} \right)^{-2/3} 
    dx^2 + \left( {\tau \over \tau_0} \right)^{4/3} \left( dy^2
    + dz^2 \right) .
     \label{KasnerSing}
\end{equation}
where $\tau_0 = 4M/3$ and $\tau = (2T^3/M)^{1/2}/3$.  In a similar
fashion, the metric may be
approximated by a flat Kasner $\left( p_1 = 1, p_2 = p_3 = 0 \right)$
solution near $T = 2M$.  There the cosmological proper time has the
asymptotic form $\tau = 4M(1-T/2M)^{1/2}$, and the asymptotic form of
the metric is
\begin{equation}
    ds^2 = - d\tau^2 + {\tau^2 \over {16 M^2}} dX^2 + 
    \left( dy^2 + dz^2 \right) \qquad ,
     \label{KasnerHorz}
\end{equation}
as $\tau \rightarrow 0$.  The singular behavior of
Eq.(\ref{KasnerHorz}) is of course only apparent; the surface $\tau =
0$ is actually the black hole event horizon.

\section{APPROXIMATE STRESS-ENERGY TENSORS}

\subsection{Massless fields}

To calculate the linearized metric perturbations to the
Schwarzschild geometry resulting from the presence of quantized
fields, it is necessary to know the values of the stress-energy
tensors of those fields.  Calculating the stress-energy tensor for a
quantized field on a black hole background spacetime is an arduous
task, which has been carried to completion only for a few cases.
Howard and Candelas have computed the stress-energy of a conformally
invariant scalar field in the Schwarzschild geometry \cite{HC,H}.
Jensen and Ottewill have computed the vacuum stress-energy of a
massless vector field in Schwarzschild \cite{JO}.  More recently
Anderson, Hiscock, and Samuel have developed a method for computing
the vacuum stress-energy for a general (arbitrary curvature coupling
and mass) scalar field in an arbitrary static spherical spacetime
and have applied their method to the
Reissner-Nordstr\"{o}m geometry\cite{AHS,AHL}.

In each of these studies, an analytic expression for $\langle T_{\mu
\nu} \rangle$ has been developed as a consequence of the procedure
used to compute the exact values for $\langle T_{\mu \nu} \rangle$.
These approximate expressions are generated by using a fourth order
WKB expansion for the field modes in the unrenormalized expression for
$\langle T_{\mu \nu} \rangle$ and then subtracting off the
DeWitt-Schwinger counterterms \cite{C} to renormalize the
stress-tensor.  The resulting analytic expressions are closely related
to approximate expressions for the vacuum stress-energy derived by
Page, Brown, and Ottewill (PBO) \cite{P,BO,PBO} and Frolov and
Zel'nikov (FZ) \cite{FZ1}.  The analytic approximation found by Howard
and Candelas is identical to the PBO approximation for the conformal
scalar field's stress-energy in Schwarzschild spacetime; further, their
numerical results show that the approximation is quite accurate for
all values of $r$ down to the horizon.  In the case of the vector
field, the analytic expression derived by Jensen and Ottewill is equal
to the PBO approximation for a conformal vector field plus a traceless
term proportional to $r^{-4}$; the resulting expression yields a good
match to the numerical results for the vector field \cite{JO}.  The
analytic approximation developed by Anderson, Hiscock, and Samuel
reduces to the FZ approximation when restricted to conformal coupling;
it has generally been shown to be valid for arbitrary curvature
coupling when compared to numerical results in the
Reissner-Nordstr\"{o}m geometry (which, of course, includes
Schwarzschild as a special case).

Each of these expressions has been derived in the exterior region of
the black hole.  There is good reason to believe they are valid in the
interior also.  The components of the curvature tensors in an
orthonormal frame are analytic functions of $r$ near the event
horizon.  Each of the approximations is also an analytic function of
the radial coordinate $r$ near the event horizon.  Thus the analytic
extension of these approximations into the interior region is trivial
to obtain.  Further Candelas and Jensen\cite{CJ} have numerically
computed $<\phi^2>$ in the interior of a Schwarzschild black hole when
the field is in the Hartle-Hawking state.  They find that Page's
approximation\cite{P} for $<\phi^2>$ arises in a natural way from the
calculation of the renormalized Feynman Green function in the interior
region and that it is a good approximation in much of the interior
region.

In this paper the Anderson, Hiscock, Samuel approximate analytic
stress-energy tensor will be used to describe the effects of quantized
massless scalar fields with arbitrary curvature coupling in the
Schwarzschild interior.  The Jensen-Ottewill analytic approximation
will be used for the stress-energy tensor of massless vector fields.
Massless spinor fields will be treated using the PBO approximation.
It should be kept in mind, however, that the spinor field expression
has not yet been tested against an accurate numerical computation to
establish its validity.

The components of the stress-energy tensor in Schwarzschild
coordinates may then be expressed as follows
\begin{equation}
    \langle T_{\mu \nu} \rangle= C_{\mu \nu} + \left(\xi - 
    1/6 \right) D_{\mu \nu}   ,
    \label{SchwarzTmn}
\end{equation}
where $C_{\mu \nu}$ represents the conformally invariant contribution
to the vacuum stress-energy from all the fields, and $D_{\mu \nu}$
represents the non-conformal contribution due to the scalar fields,
which we allow to have arbitrary curvature coupling.  Applying the
approximations discussed above:
\begin{eqnarray}
    C_{T}^{T}& = &{\epsilon \over {\lambda M^2}}  \left\{ a \left[1 +
    2 \left({2M \over T}\right) + 3 \left({2M \over T}\right)^2 
    \right] + a_3 \left({2M \over T} \right)^3 \right . \nonumber \\  
    & & + \left .  a_4 \left({2M \over T}\right)^4
    + a_5 \left({2M \over T}\right)^5 + a_6 \left({2M \over T}
    \right)^6 \right\} \qquad,
     \label{CTT}
\end{eqnarray}
where
\begin{equation}
   a = h \left( 0 \right) + {7 \over 8}h \left( 1/2 \right) + h 
   \left( 1 \right)\qquad,
   \label{a}
\end{equation}
\begin{equation}
   a_3 = 4h \left( 0 \right) - {13 \over 2}h \left( 1/2 \right) 
   - 76h \left( 1 \right)\qquad,
     \label{a3}
\end{equation}
\begin{equation}
   a_4 = 5h \left( 0 \right) - {35 \over 8}h \left( 1/2 \right) 
   + 295h \left( 1 \right)\qquad,
   \label{a4}
\end{equation}
\begin{equation}
   a_5 = 6h \left( 0 \right) - {9 \over 4}h \left( 1/2 \right) 
   - 54h \left( 1 \right)\qquad,
     \label{a5}
\end{equation}
\begin{equation}
   a_6 = 15h \left( 0 \right) + {15 \over 8}h \left( 1/2 \right)
   + 285h \left( 1 \right)\qquad,
     \label{a6}
\end{equation}
\begin{eqnarray}
    C_{x}^{x}& =& {\epsilon \over {\lambda M^2}}  \left\{ -a 
    \left[1 + 2\left({2M \over T}\right) + 3\left({2M \over T}
    \right)^2 + 4\left({2M \over T}\right)^3 \right] \right . 
    \nonumber \\ & & + \left . b_4 \left({2M \over T}\right)^4
    + b_5 \left({2M \over T}\right)^5 +  b_6 \left({2M \over T}
    \right)^6 \right\} \qquad,
    \label{Cxx}
\end{eqnarray}
where
\begin{equation}
   b_4 = -5h \left( 0 \right) - {45 \over 8}h \left( 1/2 \right)
   + 105h \left( 1 \right)\qquad,
     \label{b4}
\end{equation}
\begin{equation}
   b_5 = -6h \left( 0 \right) - {31 \over 4}h \left( 1/2 \right) 
   - 26h \left( 1 \right)\qquad,
     \label{b5}
\end{equation}
\begin{equation}
   b_6 = 33h \left( 0 \right) + {161 \over 8}h \left( 1/2 \right) 
   + 83h \left( 1 \right)\qquad,
     \label{b6}
\end{equation}
and
\begin{eqnarray}
   C_{\theta}^{\theta} = C_{\phi}^{\phi} & = & {\epsilon \over
   {\lambda M^2}} \left\{  a \left[1 + 2\left({2M \over T}\right) 
   + 3 \left({2M \over T} \right)^2 \right] + c_3 \left({2M 
   \over T}\right)^3 \right . \nonumber \\ & & + \left . c_4
   \left({2M \over T}\right)^4 + c_5 \left({2M \over T}\right)^5
   + c_6 \left({2M \over T}\right)^6 \right\} \qquad,
   \label{Cpp}
\end{eqnarray}
\begin{equation}
   c_3 = 4h \left( 0 \right) + {17 \over 2}h \left( 1/2 \right) 
   + 44h \left( 1 \right)\qquad,
     \label{c3}
\end{equation}
\begin{equation}
   c_4 = 5h \left( 0 \right) + {85 \over 8}h \left( 1/2 \right) 
   - 305h \left( 1 \right)\qquad,
     \label{c4}
\end{equation}
\begin{equation}
   c_5 = 6h \left( 0 \right) + {51 \over 4}h \left( 1/2 \right) 
   + 66h \left( 1 \right)\qquad,
     \label{c5}
\end{equation}
\begin{equation}
   c_6 = -9h \left( 0 \right) + {87 \over 8}h \left( 1/2 \right) 
   - 579h \left( 1 \right)\qquad.
     \label{c6}
\end{equation}
The constants $\epsilon$ and $\lambda$ are defined by $\epsilon =
\hbar/M^2$, $\lambda = 45 \cdot 2^{13} \cdot \pi^2$, and $h(s)$ is the
number of helicity states in, respectively, the scalar, spinor, and
vector fields present.  Explicitly, $h(0)$ simply counts the number of
scalar fields present, $h(1/2)$ is equal to 2 (or 4) for each two- (or
four-) component spinor field present; $h(1)$ is equal to 2 times the
number of vector fields present.  The nonconformal contribution to the
scalar field stress-energy is given by:
\begin{equation}
    D_{T}^{T} = -60 h(0){\epsilon \over {\lambda M^2}} \left({2M \over T}
    \right)^3 \left[4 - 3 \left({2M \over T}\right)\right]\left[1
     + 2\left({2M \over T}\right) + 3 \left({2M \over T}\right)^2 
     \right] \qquad,
     \label{DTT}
\end{equation}
\begin{equation}
    D_{x}^{x} = 180 h(0){\epsilon \over {\lambda M^2}} \left({2M \over T}
    \right)^4 \left[1 + 2 \left({2M \over T}\right) - 5 \left({2M 
    \over T}\right)^2 \right] \qquad,
     \label{Dxx}
\end{equation}
\begin{equation}
    D_{\theta}^{\theta} = 120 h(0){\epsilon \over {\lambda M^2}}\left({2M 
    \over T} \right)^3 \left[1 + 2 \left({2M \over T}\right) + 3 
    \left({2M \over T} \right)^2 - 12 \left({2M \over T}\right)^3
    \right] \qquad.
     \label{Dthth}
\end{equation}
These expressions exhibit a variety of interesting behavior in the
black hole interior.  The energy density,
$\rho = -\langle T_{T}^{T}\rangle$, is negative at the horizon for the
conformally coupled scalar field and the vector field; it is positive
there, however, for the spinor field and for any scalar field with
$\xi > 1/4$.  The energy density diverges negatively as the
singularity is approached for all conformal fields; however, the
density diverges positively for scalar fields with $\xi < 5/36$, which
includes the minimally coupled scalar field.  There is a particular
surface, $T = 3M/2$, on which the energy density of the scalar field
is independent of the curvature coupling.

The spatial stress in the $x$-direction, $\langle T_{x}^{x} \rangle$,
is positive at the horizon for all scalar fields with $\xi < 4/15$,
which includes both the minimally coupled and conformally coupled
cases, and for the conformal vector field.  The stress is negative at
the horizon for the spinor field.  This stress diverges in a positive
fashion as the singularity is approached for all conformal fields and
also for the minimally coupled scalar field.

The tangential stress, $\langle T_{\theta}^{\theta} \rangle$, is
everywhere positive in the domain of interest for the minimally
coupled scalar field and the spinor field; it is also everywhere
negative for the vector field.  The conformal scalar field has
$\langle T_{\theta}^{\theta} \rangle$ positive at the horizon, but
diverging negatively as the singularity is approached.

\subsection{Massive fields}

The technique of choice for computing an approximate renormalized
stress-energy tensor in the massive case is the DeWitt-Schwinger
approximation for $\langle T_{\mu}^{\nu} \rangle$.  This is obtained
by performing the DeWitt-Schwinger expansion of the stress-energy
tensor, in inverse square powers of the field mass, $m$,  and then
subtracting off the first, divergent terms of the expansion
\cite{dtgf}.  The remaining terms of the asymptotic series may be used
as an analytic approximation to $\langle T_{\mu}^{\nu} \rangle$.  In
this paper, approximations for the stress-energy tensor of massive
quantized fields have been derived from the previous work of Frolov
and Zel'nikov \cite{FZ84}, who used the DeWitt-Schwinger approximation
to find the renormalized stress-energy for massive fields in the Kerr
spacetime.  For the massive scalar field in the Schwarzschild limit,
Frolov and Zel'nikov's Kerr results have been found to reduce to the
stress-energy obtained by other renormalization methods \cite{FZ82}.

By taking the zero angular momentum limit ($a \rightarrow 0$) of the
Kerr results, the DeWitt-Schwinger approximation to the stress-energy
in Schwarzschild may be found for an arbitrary collection of scalar,
spinor, and vector fields.  The resulting stress-energy tensor may
again be decomposed into the contributions of the conformally
invariant fields, $C_{\mu}^{\nu}$, and the contribution of a possibly
nonconformal scalar field, $D_{\mu}^{\nu}$, according to
Eq.(\ref{SchwarzTmn}).  The components of the approximate
stress-energy tensor for conformally coupled massive fields are:

\begin{eqnarray}
C_{T}^{T}&= &{M^2 \over {1440\pi^2 T^8}}\left\{\left[15-11
	\left({{2M} \over T}\right)\right]{1 \over {m_{0}^2}}
	+\left[36-28\left({{2M} \over T}\right)\right]{1
	\over {m_{1/2}^2}} \right . \nonumber \\ & & \left .
	+\left[-99+75\left({{2M} 
	\over T}\right)\right]{1 \over {m_{1}^2}}
	\right\} \qquad ,
	\label{CTTDS}  
\end{eqnarray}
\begin{eqnarray}
C_{x}^{x}& = & {M^2 \over {10080\pi^2 T^8}}\left\{\left[-285+ 313
	\left({{2M} \over T}\right)\right]{1 \over {m_{0}^2}}
	+ \left[-540+596\left({{2M} \over T}\right)\right]
	{1 \over {m_{1/2}^2}} \right . \nonumber \\ & & \left . +
	\left[1665-1833\left({{2M} \over T}\right)\right]
	{1 \over {m_{1}^2}}\right\} \qquad ,
\label{CxxDS}
\end{eqnarray}
\begin{eqnarray}
C_{\theta}^{\theta} & = & {M^2 \over {10800\pi^2 T^8}}\left\{\left[
	-315+367\left({{2M} \over T}\right)\right]{1 \over
	{m_{0}^2}}+\left[-756+884\left({{2M} \over T}\right)\right]
	{1 \over {m_{1/2}^2}} \right . \nonumber \\ & & \left .
	+\left[2079-2427
	\left({{2M} \over T}\right)\right]{1 \over {m_{1}^2}}
	\right\},
\label{CththDS}
\end{eqnarray}
where $m_{0}$, $m_{1/2}$, and $m_{1}$ are the ``effective masses'' of
the scalar, Dirac spinor, and vector fields present. If there is no
field present for a particular spin, then its effective mass is 
set equal to infinity. If there are multiple fields with a given
spin, possibly with differing masses ({\it e.g.}, the massive 
spin $1/2$ fields in nature, representing the differing leptons
and quarks), then the effective mass is calculated according to:
\begin{equation}
	{1 \over m_{eff}^{2}} = \sum_{i=1}^{n} {1 \over m_{i}^{2}},
	\label{emass}
\end{equation}
where the sum on the right hand side is taken over the $n$ fields
of given spin present.

The nonconformal scalar stress-energy contribution is given by
\begin{equation}
D_{T}^{T} = {M^2 \over {20\pi^2 m_{0}^{2} T^8}}\left[-4+3
	\left({{2M} \over T}\right)\right] ,
\label{DTTDS}
\end{equation}
\begin{equation}
D_{x}^{x} = {M^2 \over {20\pi^2 m_{0}^{2} T^8}}\left[10-11
	\left({{2M} \over T}\right)\right] ,
\label{DxxDS}
\end{equation}
\begin{equation}
D_{\theta}^{\theta} = {M^2 \over {10\pi^2 m_{0}^{2} T^8}}
	\left[6-7\left({{2M} \over T}\right)\right] .
\label{DththDS}
\end{equation}

The DeWitt-Schwinger approximation for the stress-energy will be valid
for sufficiently massive fields, when the Compton wavelength of the
field, $\lambdabar = \hbar/m$, is much smaller than the horizon radius
of the black hole.

As was the case with the massless fields, these expressions show
interesting behavior in the interior of the black hole.  At the
horizon, the energy density, $\rho = - \langle T_T^T \rangle$, is
negative for all scalar fields with $\xi < 2/9$, which includes the
conformally and minimally coupled scalar fields.  The spinor field has
negative energy density at the horizon as well, whereas the vector
field has positive energy density.  As the singularity is approached
the energy density diverges in a positive fashion for scalar fields
with $\xi < 47/216$, which again includes both the conformally and
minimally coupled scalar fields.  The energy density of the spinor
field has a similar positive divergence, while the vector field energy
density diverges negatively.  Just as in the massless field case, the
energy density of the scalar field is independent of the curvature
coupling on the surface $T = 3M/2$.

The spatial stress in the $x$-direction, $\langle T_x^x \rangle$, is
positive on the horizon for all scalar fields with $\xi < 2/9$,
including the minimal and conformally coupled cases.  As the
singularity is approached, the stress shows a positive divergence for
all scalar fields with $\xi < 1237/5544$.  For the spinor field, the
spatial stress is also positive on the horizon and diverges in a
positive direction as the singularity is approached.  The vector field
has negative stress in both limits.

The tangential stress, $\langle T_{\theta}^{\theta} \rangle$, is
positive for all scalar fields with $\xi < 55/252$, including the
conformal scalar field.  Again in this case, the stress for the spinor
field is positive on the horizon and as the singularity is approached,
and the vector field has negative stress in both cases.

\section{SEMICLASSICAL BLACK HOLE INTERIORS}

The linearized perturbations to the Schwarzschild metric resulting
from the stress-energy of a quantized field (within the various
analytic approximation schemes discussed in the previous section) have
been described for the massless conformal scalar field by York
\cite{Y}, for the massless vector field by Hochberg and Kephart
\cite{HK}, and for the massless spinor field by Hochberg, Kephart and
York \cite{HKY}.  The perturbed geometry associated with a quantized
massless scalar field with arbitrary curvature coupling has been
analyzed by Anderson, Hiscock, Whitesell, and York \cite{AHWY}.  In
these previous calculations it was most convenient to describe the
metric perturbations in ingoing Eddington-Finkelstein coordinates,
$\left(v, r, \theta, \phi \right)$.

The study of the interior semiclassical effects proceeds most
naturally however in terms of the original Schwarzschild coordinates
(albeit with new names in the interior).  In those coordinates, the
perturbed metric may be written in the form
\begin{equation}
    ds^2 = - \left( {2M \over T} - 1 \right)^{-1} \left[1 + 
    \epsilon \eta (T)\right] dT^2 +\left( {2M \over T} - 1 \right)
    \left[1 + \epsilon \sigma (T)\right] dx^2 + T^2 d{\Omega}^2 .
     \label{MetricCosmo}
\end{equation}
The Einstein equations, to first order in $\epsilon$, then have the
form:
\begin{equation}
    {d \over {dT}} \left[\left(2M - T \right)\eta\right] = { {8 \pi
    T^2 \langle T_{x}^{x}\rangle} \over \epsilon}\qquad,
     \label{EtaEEqn}
\end{equation}
\begin{equation}
    {{d \sigma} \over {dT}}  = -{ {8 \pi T^2 \langle T_{T}^{T}\rangle}
    \over 
    {\epsilon\left(2 M - T \right)}} - {\eta \over {2 M - T}}
    \qquad.
     \label{SigEEqn}
\end{equation}

\subsection{Massless fields}

Integrating Eqs.(\ref{EtaEEqn},\ref{SigEEqn}) using the approximate
stress-energy tensor for a collection of massless quantized fields given
in Eqs.(\ref{CTT}-\ref{Dthth}), one obtains
\begin{eqnarray}
    K \eta& =& A \left[ \left({T \over {2M}}\right)^2 + 4\left(
    { T\over {2M}}
    \right) + 12 \left(1 - {T \over {2M} }\right)^{-1} 
    \ln\left({{2M} \over T} \right) \right]  \nonumber
    \\ & & +A_0
    + A_1 \left({{2M} \over T} \right) 
    + A_2 \left({{2M} \over T} \right)^2
    + A_3 \left({{2M} \over T} \right)^3 \qquad,
     \label{EtaSoln}
\end{eqnarray}
\begin{eqnarray}
    K \sigma& =& A \left[ \left({T \over {2M}}\right)^2 + 8 \left(
    {T \over
    {2M}} \right) - 24 \left({{3M - T} \over {2M - T}}\right) 
    \ln\left( {2M} \over T \right) \right] \nonumber
    \\ & & + B_0 + 
    B_1 \left({{2M} \over T} \right) + B_2 \left({{2M} \over T} \right)^2 
    + B_3 \left({{2M} \over T} \right)^3 \qquad,
     \label{SigSoln}
\end{eqnarray}
where $K = 3840 \pi$, and the coefficients $A_i$, $B_i$ are given by
\begin{equation}
    A = { {8 h \left( 0 \right) + 7 h \left( 1/2 \right) + 8 h 
    \left( 1 \right)} \over 24 }\qquad,
     \label{A}
\end{equation}
\begin{equation}
	A_0 = {1 \over 24}\left[8\left(109 - 360 \xi \right)
	h \left( 0 \right) + 43 h \left( 1/2 \right)
	+375 h \left( 1 \right) \right] \qquad,
\label{A0}
\end{equation}
\begin{equation}
    A_1 = {1 \over 24}\left[ 8 \left(1 - 60 \xi \right) h 
    \left( 0 \right)  + 67 h \left( 1/2 \right) - 2872 h \left( 
    1 \right) \right]\qquad,
     \label{A1}
\end{equation}
\begin{equation}
    A_2 = {1 \over 6} \left[ 8 \left(-11 + 30 \xi \right) 
    h \left( 0 \right) - 17 h \left( 1/2 \right) - 88 h \left( 
    1 \right) \right] \qquad,
     \label{A2}
\end{equation}
\begin{equation}
    A_3 = {1 \over 24} \left[ 8 \left(-83 + 300 \xi \right) h 
    \left( 0 \right) - 161 h \left( 1/2 \right) - 664 h \left( 1 
    \right) \right] \qquad,
     \label{A3}
\end{equation}
\begin{equation}
    B_0 = {1 \over 24} \left[ 8 \left(155 - 720 \xi \right) h 
    \left( 0 \right) + 365 h \left( 1/2 \right) - 565 h \left( 
    1 \right) \right] + k_0\qquad,
     \label{B0}
\end{equation}
\begin{equation}
    B_1 = {1 \over 8} \left[ 8 \left(-27 + 100 \xi \right) h 
    \left( 0 \right) - 89 h \left( 1/2 \right) + 1064 h \left( 
    1 \right) \right]\qquad,
     \label{B1}
\end{equation}
\begin{equation}
    B_2 = {1 \over 12} \left[ 8 \left(-23 + 120 \xi \right) 
    h \left( 0 \right) - 41 h \left( 1/2 \right) + 296 h \left
    ( 1 \right) \right]\qquad,
     \label{B2}
\end{equation}
and
\begin{equation}
    B_3 = { 5 \over 24} \left[ 8 \left(-5 + 36 \xi \right) 
    h \left( 0 \right) + h \left( 1/2 \right) + 152 h \left( 1 
    \right) \right]\qquad.
     \label{B3}
\end{equation}

The form of Eq.  (\ref{B0}) has been chosen so that the integration
constant in $\sigma$ is expressed in terms of the integration
constant, $k_0$, which has appeared in previous papers\footnote{In
each of these papers the black hole was surrounded by a thin perfectly
reflecting cavity.  The specific value of the integration constant
$k_0$ was obtained in those cases by requiring $g_{tt}$ to be 
continuous at the cavity wall. In the present work, none of our
results will depend on the  numerical value chosen for $k_0$.} \cite{Y},
\cite{HK}, \cite{HKY}, \cite{AHWY}.  The integration constant which is
associated with $\eta$ has been absorbed via renormalization into $M$;
the constant $M$ which appears in these equations is thus to be
interpreted as the ``dressed'' mass of the black hole.

The semiclassical metric of Eq.  (\ref{MetricCosmo}) is valid only
when the perturbations, $\epsilon \eta$ and $\epsilon \sigma$, are
small compared to unity.  The perturbations are small at the horizon,
$T = 2 M$, for black hole masses greater than or equal to the Planck
mass (recall $\epsilon = \hbar/M^2 = M_{P}^{2}/M^2$).  Of course, the
perturbations can always be made large by taking the large-N limit,
where N is the number of quantized fields present.  For reasonable
numbers of fields, and black hole masses greater than the Planck mass,
it is possible to approach the singularity at $T = 0$ fairly closely.
As an example, if we take $h(0) = 0$, $h(1/2) = 6$, $h(1) = 2$,
representing three massless neutrino fields and one massless vector
field, and a black hole mass of $M = M_P$, then the perturbations
reach a strength of $10^{-1}$ at about $T = M$; for a solar mass black
hole, however, the perturbation does not reach this strength until $T
\approx 3 \times 10^{-21} \mbox{cm} = 2 \times 10^{-26} M$.

\subsection{Massive fields}

Integrating Eqs.(\ref{EtaEEqn},\ref{SigEEqn}) using the approximate
stress-energy tensor for a collection of massive quantized fields given
in Eqs.(\ref{CTTDS}-\ref{DththDS}), one obtains
\begin{equation}
	K\eta = E\left[\left({2M \over T}\right)+\left({2M \over 
	T}\right)^2+\left({2M \over T}\right)^3+\left({2M \over 
	T}\right)^4+\left({2M \over T}\right)^5\right]+\tilde{E}
	\left({2M \over T}\right)^6 ,
\label{etads}
\end{equation}
\begin{eqnarray}
	K\sigma& =& k_0-E\left[-5+\left({2M \over T}\right)+\left({2M 
	\over T}\right)^2+\left({2M \over T}\right)^3+\left({2M 
	\over T}\right)^4+\left({2M \over T}\right)^5\right] \nonumber
	\\ & & +F\left[\left({2M \over T}\right)^6-1\right]  ,
\label{sigds}
\end{eqnarray}
where $K$ is again equal to $3840\pi$, and
\begin{equation}
	E={1 \over {126 M^2}}\left[\left(113-504\xi\right)
	{1 \over {m_{0}^{2}}}+52{1 \over {m_{1/2}^{2}}}-165
	{1 \over {m_{1}^{2}}}\right],
\label{Eeta}
\end{equation}
\begin{equation}
	\tilde{E}={1 \over {126 M^2}}\left[\left(-1237
	+5544\xi\right) {1 \over {m_{0}^{2}}}-596{1 \over 
	{m_{1/2}^{2}}}+1833{1 \over {m_{1}^{2}}}\right],
\label{Eteta}
\end{equation}
\begin{equation}
	F={1 \over {18 M^2}}\left[\left(-47+216\xi\right)
	{1 \over {m_{0}^{2}}}-28{1 \over {m_{1/2}^{2}}}+75
	{1 \over {m_{1}^{2}}}\right].
\label{Fsig}
\end{equation}
The integration constants in Eqs.(\ref{etads},\ref{sigds}) are
handled in the same manner as in the massless case; in particular,
the black hole mass $M$ is the ``dressed'' or renormalized mass.
The field masses $m_0$, $m_{1/2}$, $m_{1}$, are effective masses
defined as described in Sec. III.

The perturbations of the Schwarzschild metric caused by the 
presence of massive fields are small, and the DeWitt-Schwinger
approximation valid, so long as the Compton wavelength of the
field is significantly less than the local radius of curvature
of the spacetime. In the Schwarzschild interior, this will be
true so long as $T >> (M/m^2)^{1/3}$.

\section{ANISOTROPY OF THE SCHWARZSCHILD INTERIOR}

Since the Schwarzschild interior represents a highly anisotropic
cosmology, it is natural to ask whether semiclassical effects dampen
or strengthen the anisotropy.  Many studies over the last quarter
century have established that particle production can rapidly
isotropize an anisotropic cosmology
\cite{Z,ZS,H1,HFP,FPH,BB,LS,HP,HH}.  As mentioned in the introduction,
the analytical approximations for massless fields are nonlocal
and thus probably take
particle production into account to some extent.  However, it is
completely unknown at this point how well they do this.  The
DeWitt-Schwinger approximation for the massive fields does not take
particle production into account at all because it is a local
approximation and particle production is an intrinsically nonlocal
phenomenon.  Thus whatever dissipation of anisotropy that is found due
to all of these approximations is likely to be less that what would
occur if full numerical solutions to the nonlinear backreaction
equations were obtained.

One measure of the anisotropy of the interior is the ratio of the
Hubble expansion rates in the differing spatial directions.  In the
present case, since the two spatial directions on the two-spheres of
symmetry are equivalent, there is only one ratio to calculate, say
\begin{equation}
    \alpha = { H_x \over H_\theta } = { {g_{\theta \theta} { {d 
    g_{x x}} \over {d \tau}}} \over {g_{x x} { {d g_{\theta \theta}}
    \over {d \tau}}} } = { {g_{\theta \theta} { {d g_{x x}} \over
    {dT}}} \over {g_{x x} { {d g_{\theta \theta}} \over {dT}}} } 
    \qquad.
     \label{Aniso}
\end{equation}

The sign of $\alpha$ is positive if the cosmology is expanding (or
contracting) in all three spatial directions.  If the cosmology is
expanding or contracting isotropically, then $\alpha = 1$.

Evaluating $\alpha$ for the metric of Eq.  (\ref{MetricCosmo}), we
find, to first order in $\epsilon$
\begin{equation}
    \alpha = \alpha_{Sch} + \epsilon \delta \alpha \qquad,
     \label{AnisoExpnd}
\end{equation}
where $\alpha_{Sch}$ is the ordinary Schwarzschild value,
\begin{equation}
    \alpha _{Sch} = { -M \over {2 M - T} } \qquad,
     \label{AnisoSch}
\end{equation}
and
\begin{equation}
    \delta \alpha = {1 \over 2} T { {d \sigma} \over {dT}} \qquad.
     \label{AnisoPert}
\end{equation}

Taking Eq.  (\ref{SigEEqn}) with Eq.  (\ref{AnisoPert}), the
perturbation to the anisotropy can be written explicitly in terms of
components of the stress-energy as
\begin{equation}
    \delta \alpha = -{1 \over 2} T \left[ {\eta \over \left( 2M 
    - T \right)} + { {8 \pi T^2  \langle T_T^T\rangle } \over 
    {\epsilon \left(
    2M - T \right) }} \right] \qquad.
     \label{FullPert}
\end{equation}
If the overall sign of the perturbation to the anisotropy is positive,
then the semiclassical effects tend to isotropize the interior.
Negative values of $\delta \alpha$ push the spacetime towards greater
anisotropy.

Since the anisotropy is the ratio of the expansion rates along
different spatial directions, careful consideration must be given to
the method of spacetime slicing used to compare the perturbed and
unperturbed spacetimes.  One choice would be to consider slices which
sit at equal proper times away from the horizon.  Another choice, used
in this paper, is to consider surfaces with equal values of the
Schwarzschild area coordinate $T$.

Taking the stress-energy tensors described in the previous section for
the quantized fields of interest, the contributions described in Eq.
(\ref{FullPert}) can then be computed for various spin fields on the
Scwarzschild background.  It should be noted when considering these
results that the perturbation expansions become less reliable as one
proceeds away from the horizon and towards the singularity, but the
exact point at which the perturbation should no longer be trusted is a
matter of choice.

The perturbation to the anisotropy in the presence of a 
massless scalar field is
\begin{eqnarray}
    \delta \alpha & = & {1 \over {\pi \left( 2M - T \right)^2}} 
    \left\{ M^2 \left[ {\xi \over 48} - {17 \over 2880} \right]
    + {M^3 \over T} \left[ {\xi \over 24} - {7 \over 720} \right]
    + {M^4 \over T^2} \left[{ {5 \xi} \over 12} - {29 \over 720} 
    \right] \right. \nonumber \\ & & + {M^5 \over T^3} \left[ 
    {5 \over 48} - { {3 \xi} \over 4} \right]
    + MT  \left[ {29 \over 11520} - { {5 \xi} \over 192} - { {\ln
    \left( 2M/T \right)} \over 960} \right] \nonumber \\ & & 
    + \left. {T^2 \over 2304} + {T^3 \over {11520 M}} + {T^4 
    \over {46080 M^2}} \right\} \qquad .
    \label{ScalarMsLs}
\end{eqnarray}
The sign of $\delta \alpha$ clearly depends on the value of the scalar
curvature coupling, $\xi$.  For values of $\xi < 5/36$ the
perturbation is positive, and the field tends to isotropize the
spacetime.  For values of $\xi > 12/55$ the perturbation is negative
and the spacetime tends to more anisotropy.  Between these two values,
$5/36 < \xi < 12/55$, the perturbation isotropizes in some regions of
the interior and anisotropizes in other regions, as shown in Figure 1.
For values of $\xi$ and $T$ above the solid line, the spacetime is
pushed towards anisotropy.  Values of $\xi$ and $T$ below the solid
line make $\delta \alpha > 0$, and the spacetime is isotropized in the
presence of the scalar field.  In this case, the minimally coupled
scalar field ($\xi = 0$) always isotropizes the spacetime, whereas the
conformally coupled field ($\xi = 1/6$) only isotropizes in the
interior regions near the horizon.

The perturbation due to the massless spin-$1/2$ field is
\begin{eqnarray}
    \delta \alpha &=& {1 \over {\pi \left( 2M - T \right)^2}} 
    \left\{ - {M^5 \over T^3} {1 \over 192} + { {M^4 \over T^2}
    {97 \over 2880}} - { {M^3 \over T}{19 \over 2880}} -
    {M^2 {59 \over 11520}} \right. \nonumber \\
    & & \left. - MT \left[ {97 \over 46080} + { {7 \ln \left(
    2M/T \right)} \over 3840} \right] + { {7 T^2} \over 9216} 
    + { {7 T^3} \over {46080 M} } + { {7 T^4} \over {184320 M^2}
    } \right\} \qquad .
    \label{SpinorMsLs}
\end{eqnarray}
For $T > 0.5$, which is the region where the perturbation expansion
can be trusted, this always pushes the spacetime towards greater
isotropy.  

The massless vector field perturbation to the anisotropy is
\begin{eqnarray}
    \delta \alpha & = & {1 \over {\pi \left( 2M - T \right)^2}}
    \left\{ - { {19 M^5} \over {24 T^3}} + { {211 M^4} \over 
    {360 T^2}} - { {97 M^3} \over {360 T}} + { {343 M^2} \over 
    1440} \right. \nonumber \\ & & \left. - MT \left[ {451 \over 
    5760} + { { \ln \left( 2M/T \right)} \over 480} \right] + { T^2
    \over 1152} + { T^3 \over {5760 M} } + { { T^4 \over {23040 M^2}
    }} \right\}
    \label{VectorMsLs}
\end{eqnarray}
and pushes towards anisotropy for all values of $T$ in the interior.

The impact of massive fields of varying spin can be considered as
well.  For the massive scalar field
\begin{eqnarray}
   \delta \alpha & = & { 1 \over {\pi m^2} } \left\{ {M^4 
   \over T^6} \left[ {47 \over 360} - { {3 \xi} \over 5} \right] 
   + {M^3 \over T^5} \left[ {113 \over 6048} - {\xi \over 12} \right] 
   + {M^2 \over T^4} \left[ {113 \over 15120} - {\xi \over 240} 
   \right]\right. \nonumber \\ & & \left. + {M \over T^3} \left[ {113 
   \over 40320} - {\xi \over 80} \right] + {1 \over T^2} \left[ {113 
   \over 120960} - {\xi \over 240} \right] + {1 \over {MT} } \left[ 
   {113 \over 483840} - {\xi \over 960} \right] \right\}
   \label{ScalarMsve}
\end{eqnarray}
where $m$ is the effective field mass defined in Eq.(\ref{emass}).
Similar to the case of the massless scalar field, the exact sign of
the perturbation depends on the value of the scalar curvature
coupling.  When $\xi > 1223/5544$, the presence of the field makes the
spacetime more anisotropic, and when $\xi < 47/216$ the push is always
towards isotropy.  As shown in Figure 2, there exists a range of
values $47/216 < \xi < 1223/5544$ over which some interior regions are
isotropized and others are not.  As before, values of $\xi$ and $T$
above the solid line have $\delta \alpha < 0$ and the spacetime tends
towards anisotropy.  For values below the solid line, $\delta \alpha >
0$, and the tendency is towards isotropy.  Both minimal and conformal
coupling fall within this regime.

The perturbation due to a massive spinor field is
\begin{equation}
   \delta \alpha = { 1 \over {\pi m^2} } \left\{ {M^4 \over T^6}
   {7 \over 90} + {M^3 \over T^5}{113 \over 1512} + {M^2 \over T^4}
   {13 \over 3780} + {M \over T^3}{13 \over 10080} + {1 \over T^2}
   {13 \over 30240} + {1 \over {MT} }{13 \over 120960} \right\}
   \label{SpinorMsve}
\end{equation}
which is manifestly positive for all values of $T$, and hence 
decreases the anisotropy.

Similarly, the massive vector field perturbation to the anisotropy is
\begin{equation}
   \delta \alpha = -{ 1 \over {\pi m^2} } \left\{ {M^4 \over T^6}
   {5 \over 24} + {M^3 \over T^5}{55 \over 2016} + {M^2 \over T^4}
   {11 \over 1008} + {M \over T^3}{11 \over 2688} + {1 \over T^2}
   {11 \over 8064} + {1 \over {MT} }{11 \over 32256} \right\}
   \label{VectorMsve}
\end{equation}
which is manifestly negative for all $T$, and so always tends to
increase the anisotropy.

\section{APPROACHING THE FINAL SINGULARITY}

Ever since it was realized that singularities could not be avoided in
physically plausible spacetimes, quantum effects have been invoked as
the physical instrument which might restore regularity to spacetime,
by banishing singular behavior.  While it is impossible for a
perturbative analysis to determine whether quantum effects will
eradicate the singularity, it is possible, it is possible to determine
how the growth of curvature as one approaches the singularity is
affected by the semiclassical perturbation.

The simplest way to see the effect of quantized fields on the growth
of curvature as one approaches the singularity is to examine the
perturbations of curvature scalars.  One such scalar is the
Kretschmann scalar, which for unperturbed Schwarzschild is
\begin{equation}
    K_{Sch} = R^{\alpha \beta \mu \nu} R_{\alpha \beta \mu \nu} =
    { {48 M^2} \over T^6} - { {32 M} \over T^5} + {16 \over T^4} 
    \qquad .
    \label{KScalar}
\end{equation}
The Kretschmann scalar is perfectly well behaved near the
horizon $T = 2M$, but diverges strongly as $T \rightarrow 0$.

Evaluating $K$ to first order in $\epsilon$ for the metric of Eq. 
(\ref{MetricCosmo}) will yield
\begin{equation}
   K = K_{Sch} + \epsilon \delta K \qquad .
   \label {KExpand}
\end{equation}
The first order correction to the Kretschmann scalar can be written in
terms of the perturbation functions $\eta$ and $\sigma$ as
\begin{eqnarray}
   \delta K& =& {8 \over T^3} \left[ -12 \eta {M^2 \over T^3} + 6 \eta 
   {M \over T^2} - 2 \eta {1 \over T} + 3 \eta^\prime {M^2 \over T^2} 
   - \eta^\prime {M \over T} \right . \nonumber \\ & & \left .
   - 5 \sigma^\prime {M^2 \over T^2} 
   + \sigma^\prime {M \over T} + 2 \sigma^{\prime\prime} {M^2 \over T}
   - \sigma^{\prime\prime} M \right]
   \label {DeltaK}
\end{eqnarray}
where primes denote differentiation with respect to $T$.

If the sign of $\delta K$ is positive, the divergence as one
approaches the curvature singularity will be strengthened.  If $\delta
K$ is negative, then the divergence will be weakened.

The perturbation to $K_{Sch}$ in the presence of a massless scalar 
field is
\begin{eqnarray}
    \delta K & = & {{128 \pi} \over {\lambda T^2}} \left\{ {M^5 \over
    T^7} \left[ 12288 + 11520 \xi \right] - {M^4 \over T^6} 
    \left[ 7712 - 24960 \xi \right] - {M^3 \over T^5} \left[
    192 + 2880 \xi  \right] \right. \nonumber \\ & & +
    {M^2 \over T^4} \left[ 816 - 7200 \xi - 288 \ln \left( 2M/T \right)
    \right] + {M \over T^3} \left[ 168 + 480 \xi + 96 \ln \left( 2M/T 
    \right) \right] \nonumber \\ & & - \left. {30 \over T^2} 
    - {6 \over MT} - {1 \over M^2} \right\} \qquad .
    \label{KScalarMsLs}
\end{eqnarray}
The sign of $\delta K$ depends on the value of the scalar curvature
coupling, $\xi$.  Figure 3 shows a plot of the curvature coupling
$\xi$ vs.  $T$ over the interior.  The solid line represents values of
$\xi$ and $T$ for which $\delta K = 0$.  For points below the solid
line, the perturbation to the Kretschmann scalar is negative, and for
values above the line the perturbation is positive.  For all
nonnegative values of the curvature coupling, the contribution is
positive, and hence curvature grows faster than in the unperturbed
metric.

The massless spinor field perturbs $K_{Sch}$ by
\begin{eqnarray}
    \delta K & = & {{224 \pi} \over {\lambda T^2}} \left\{ {57216
    \over 7} {M^5 \over T^7}  - {29024 \over 7} {M^4 \over T^6} -
    {7584 \over 7} {M^3 \over T^5}  \right. \nonumber \\ & & + {M^2
    \over T^4} \left[ {48 \over 7} - 288 \ln \left( 2M/T \right) 
    \right] + {M \over T^3} \left[ {696 \over 7} + 96 \ln \left( 
    2M/T \right) \right] \nonumber \\ & & - \left. {30 \over T^2} 
    - {6 \over MT} - {1 \over M^2} \right\} \qquad .
    \label{KSpinorMsLs}
\end{eqnarray}
The perturbation of Eq.  (\ref{KSpinorMsLs}) changes sign in the
interior, yielding a negative contribution to the Kretschmann scalar
for $T > 1.435$, and a positive contribution to the Kretschmann scalar
for $T < 1.435$.

The massless vector field perturbation to the Kretschmann scalar is
\begin{eqnarray}
    \delta K & = & {{256 \pi} \over {\lambda T^2}} \left\{ 87168
    {M^5 \over T^7}  - 15392 {M^4 \over T^6} +
    31008 {M^3 \over T^5}  \right. \nonumber \\ & & + {M^2
    \over T^4} \left[ 15024 - 288 \ln \left( 2M/T \right) \right]
    + {M \over T^3} \left[ 3048 + 96 \ln \left( 2M/T \right) \right]
    \nonumber \\ & & - \left. {30 \over T^2} - {6 \over MT} 
    - {1 \over M^2} \right\} \qquad .
    \label{KVectorMsLs}
\end{eqnarray}
which is positive for all values of $T$ in the interior; the massless
vector field thus seems to strengthen the growth of curvature as the
singularity is approached.

Similar considerations can be given to massive fields.  In the case of 
the massive scalar field
\begin{eqnarray}
    \delta K & = & {1 \over {\pi m^2 T^5}} \left\{ -{1 \over M} 
    \left[ {113 \over 15120} - {\xi \over 30} \right] + {1 \over T} 
    \left[ {113 \over 5040} - {\xi \over 10} \right] + {M^4 \over T^5} 
    \left[ {20 \over 7} - {{64 \xi} \over 5} \right] \right. \nonumber
    \\ & & - \left. {M^5 \over T^6} 
    \left[ {1076 \over 189} - {{352 \xi} \over 15} \right] - {M^6 
    \over T^7} \left[ {44 \over 105} - {{32 \xi} \over 5} \right] 
    \right\} \qquad .
    \label{KScalarMsve}
\end{eqnarray}

As in the massless scalar field case, the exact sign of the
perturbation depends on the scalar curvature coupling.  Figure 4 shows
a plot of the curvature coupling $\xi$ vs.  $T$ over the interior.
Values of $\xi$ and $T$ below the solid line yield $\delta K < 0$.
For points above the solid line, $\delta K > 0$.  The minimally
coupled massive scalar field always weakens the growth of curvature;
for the conformally coupled field, the rate of curvature growth is
initially less than in Schwarzschild, but near the singularity the
perturbation causes the curvature to grow more rapidly than in the
unperturbed metric.

For the massive spinor field
\begin{eqnarray}
    \delta K & = & {1 \over {\pi m^2 T^5}} \left\{
    -{1 \over M} {13 \over 3780}
    + {1 \over T} {13 \over 1260}
    + {M^4 \over T^5} {48 \over 35}
    - {M^5 \over T^6} {656 \over 945}
    - {M^6 \over T^7} {496 \over 105}
    \right\} \qquad 
    \label{KSpinorMsve}
\end{eqnarray}
which is negative for all interior values of the coordinate $T$;
hence the massive spinor field softens the approach to the
singularity, decreasing the rate of increase of the curvature.

In contrast, the massive vector field has
\begin{eqnarray}
    \delta K & = & {1 \over {\pi m^2 T^5}} \left\{
     {1 \over M} {11 \over 1008}
    - {1 \over T} {11 \over 336}
    - {M^4 \over T^5} {148 \over 35}
    + {M^5 \over T^6} {2012 \over 315}
    + {M^6 \over T^7} {36 \over 7}
    \right\} \qquad 
    \label{KVectorMsve}
\end{eqnarray}
which is positive for all values of $T$ in the interior; hence
the massive vector field strengthens the growth of curvature as
the singularity is approached.

\section{DISCUSSION AND SUMMARY}

In this paper we have calculated the linearized perturbations of the
Schwarzschild black hole interior due to a collection of quantized
matter fields.  The stress-energy tensor of the matter fields has been
described using analytic approximations.  For massless fields, we have
used the approximations of Page, Brown, and Ottewill \cite{PBO} for
the spinor field, the approximation of Jensen and Ottewill \cite{JO}
for the vector field, and that of Anderson, Hiscock, and Samuel
\cite{AHS} for the scalar field.  Massive fields have been treated
using the DeWitt-Schwinger approximation, as developed by Frolov and
Zel'nikov \cite{FZ84} and Anderson, Hiscock, and Samuel \cite{AHS}.

These calculations provide virtually all of the useful information
about semiclassical effects in the interior of a black hole that can
be obtained using the various analytical approximations.  One could
attempt to construct fully self-consistent solutions to the
semiclassical equations using the DeWitt-Schwinger approximation for
massive fields or the approximations of Frolov and Zel'nikov\cite{FZ1}
or Anderson, Hiscock, and Samuel\cite{AHS} for massless fields.
However, serious problems arise in such calculations.  For massless
fields the analytical approximations diverge logarithmically on the
event horizon in any static non-Ricci-flat spacetime.  Numerical
computations of the stress-energy tensor in Reissner-Nordstr\"{o}m
spacetimes\cite{AHS,AHL} indicate that these divergences are not real.
They are simply an indication that it is only for Schwarzschild
spacetime that the analytical approximations are valid near the event
horizon.  For massive fields the DeWitt-Schwinger approximation gives
no divergent behavior on the event horizon of any black hole.
However, this approximation is valid only in the limit that the mass
Compton wavelength of the field is much smaller than the radius of
curvature of the spacetime.  Thus the best that can be done when using
the DeWitt-Schwinger approximation is to solve the semiclassical
equations perturbatively, in which case the first order term is
definitely the most important.  Therefore, it will be necessary to
numerically compute the stress-energy tensor to study semiclassical
interior effects beyond the level of linear perturbation theory.

We have addressed the question of whether anisotropy is dissipated in
the interior by treating the black hole interior as an anisotropic,
homogeneous cosmology and examining whether the perturbed metric has
greater or lesser anisotropy than the background Schwarzschild metric.
We find that minimally and conformally coupled scalar fields, and the
spinor field, decrease the anisotropy as one approaches the
singularity, while vector fields increase the anisotropy.  These
results are described from the point of view of the black hole
interior, which as a cosmology is a universe approaching a final
singularity.  If one instead interpreted our results in terms of the
white hole portion of the Schwarzschild Penrose diagram, then scalar
and spinor fields would enhance anisotropy as one moves away from the
singularity, while vector fields would reduce it.  However as
previously mentioned, in this case the boundary conditions for the
fields are ``final'' rather than ``initial'' conditions.

We have also examined whether there is any evidence for the
semiclassical perturbation modifying the approach to the singularity.
While within the context of perturbation theory, it is impossible to
determine whether quantum effects might substantially change the
character of the singularity (perhaps even eliminating it), one can
ask whether an observer approaching the final black hole singularity
will measure larger or smaller curvature in the perturbed metric than
in the classical Schwarzschild case.  We find that massless fields of
all spin, and massive vector fields, generally strengthen the
singularity (curvature grows faster than in Schwarzschild) while the
massive scalar and spinor fields weaken the growth of curvature.

\acknowledgements
This work was supported in part by NSF Grants Nos. PHY92-07903 and 
PHY95-11794 (W.\ A.\ H.\ ), and PHY95-12686 (P.\ R.\ A.\ ).

\begin{figure} 
\caption{ The curve represents zero perturbation to the
anisotropy of the interior for a massless scalar field as a function
of the coordinate $T$ and the curvature coupling $\xi$.  Above the
curve, the perturbations due to the scalar field make the spacetime
more anisotropic, and below the curve they make the spacetime more
isotropic.  As $T \rightarrow 0$, $\xi$ approaches $5/36$ on the
curve.  As $T \rightarrow 2M$, $\xi$ approaches $12/55$.}
\end{figure}

\begin{figure} 
\caption{ The curve represents zero perturbation to the
anisotropy of the interior for a massive scalar field as a function of
the coordinate $T$ and the curvature coupling $\xi$.  Above the curve,
the perturbations make the spacetime more anisotropic, and below the
curve they make the spacetime more isotropic.  As $T \rightarrow 0$,
$\xi$ approaches $47/216$ on the curve.  As $T \rightarrow 2M$, $\xi$
approaches $1223/5544$.}  
\end{figure}

\begin{figure} 
\caption{ The curve represents zero perturbation to the
Kretschmann scalar for a massless scalar field as a function of the
coordinate $T$ and the curvature coupling $\xi$.  Above the curve, the
perturbations are positive, increasing the rate of curvature growth,
and below the curve the perturbations are negative, decreasing the
growth of curvature relative to the classical solution.}  
\end{figure}

\begin{figure} 
\caption{ The curve represents zero perturbation to the
Kretschmann scalar for a massive scalar field as a function of the
coordinate $T$ and the curvature coupling $\xi$.  Above the curve, the
perturbations to Kretschmann are positive, and below the curve the
perturbations are negative.}  \end{figure}

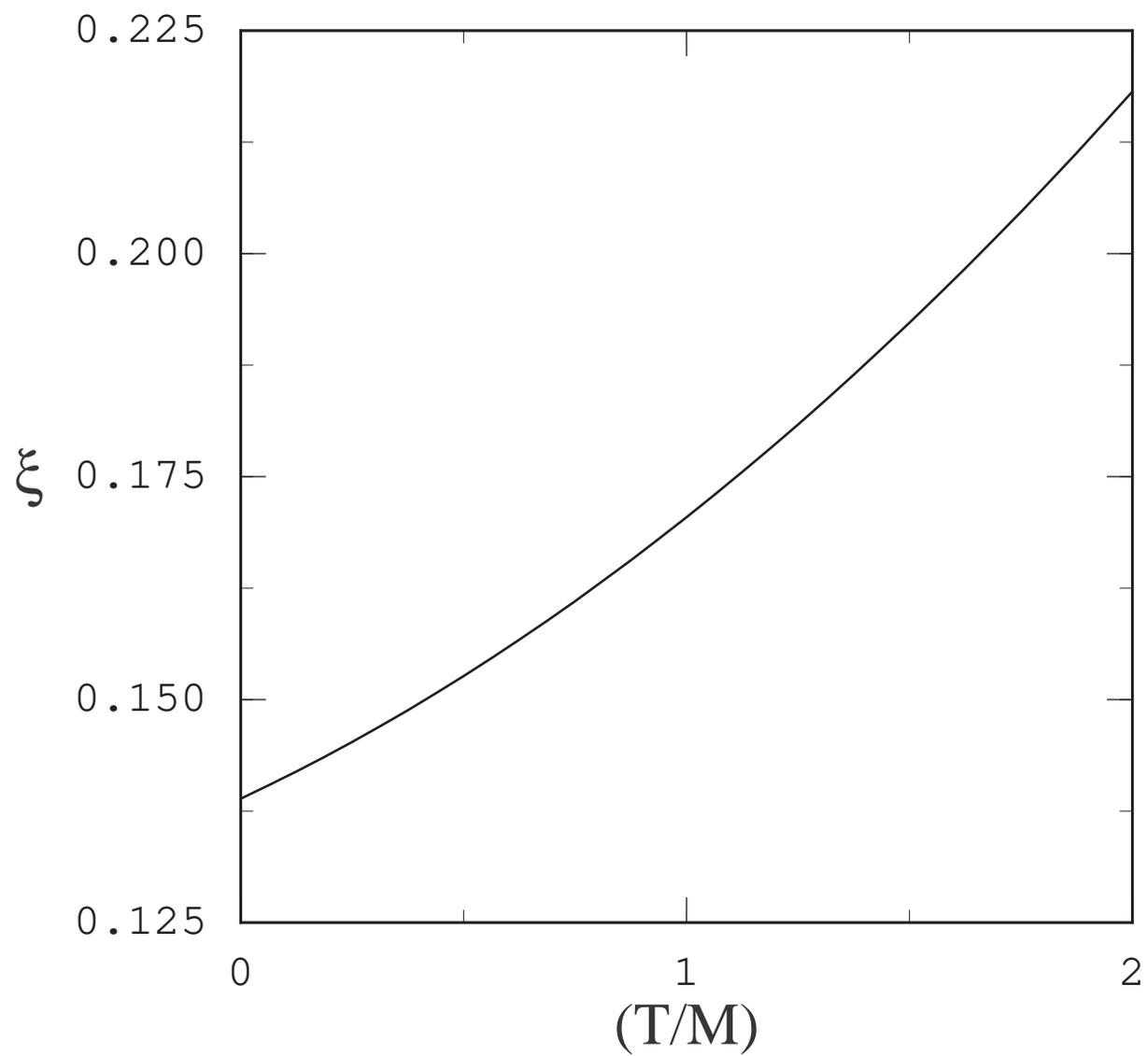
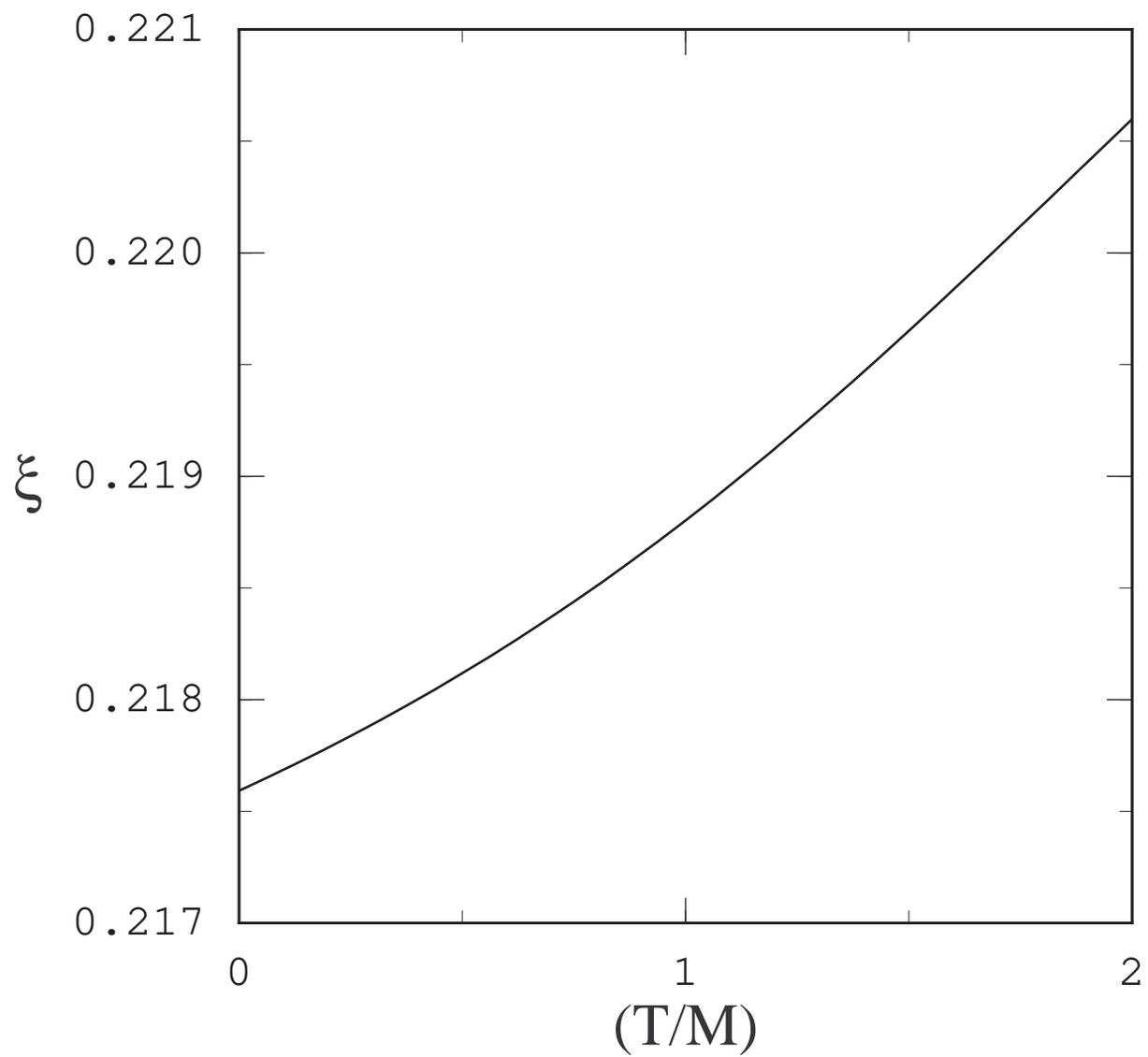
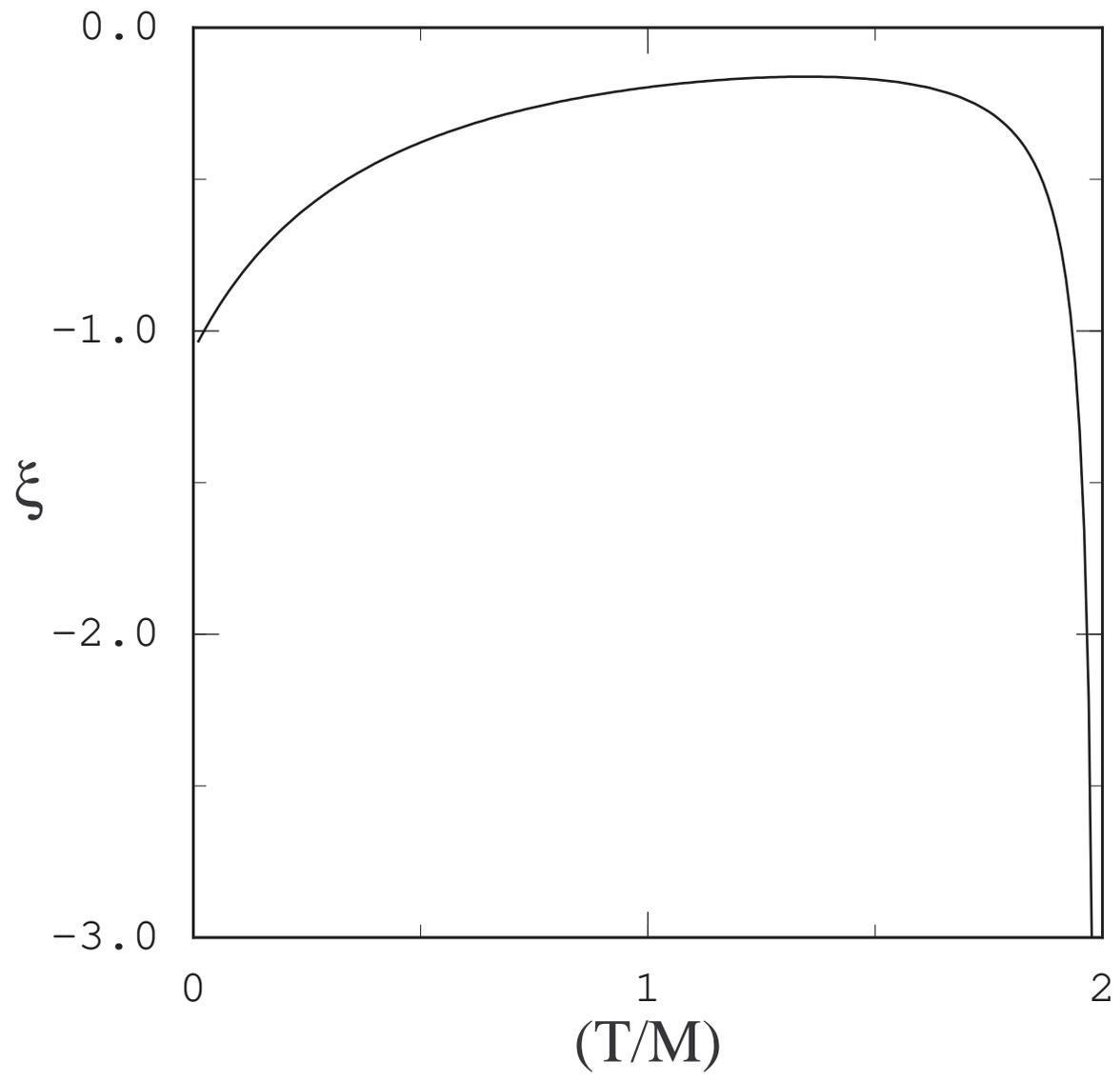
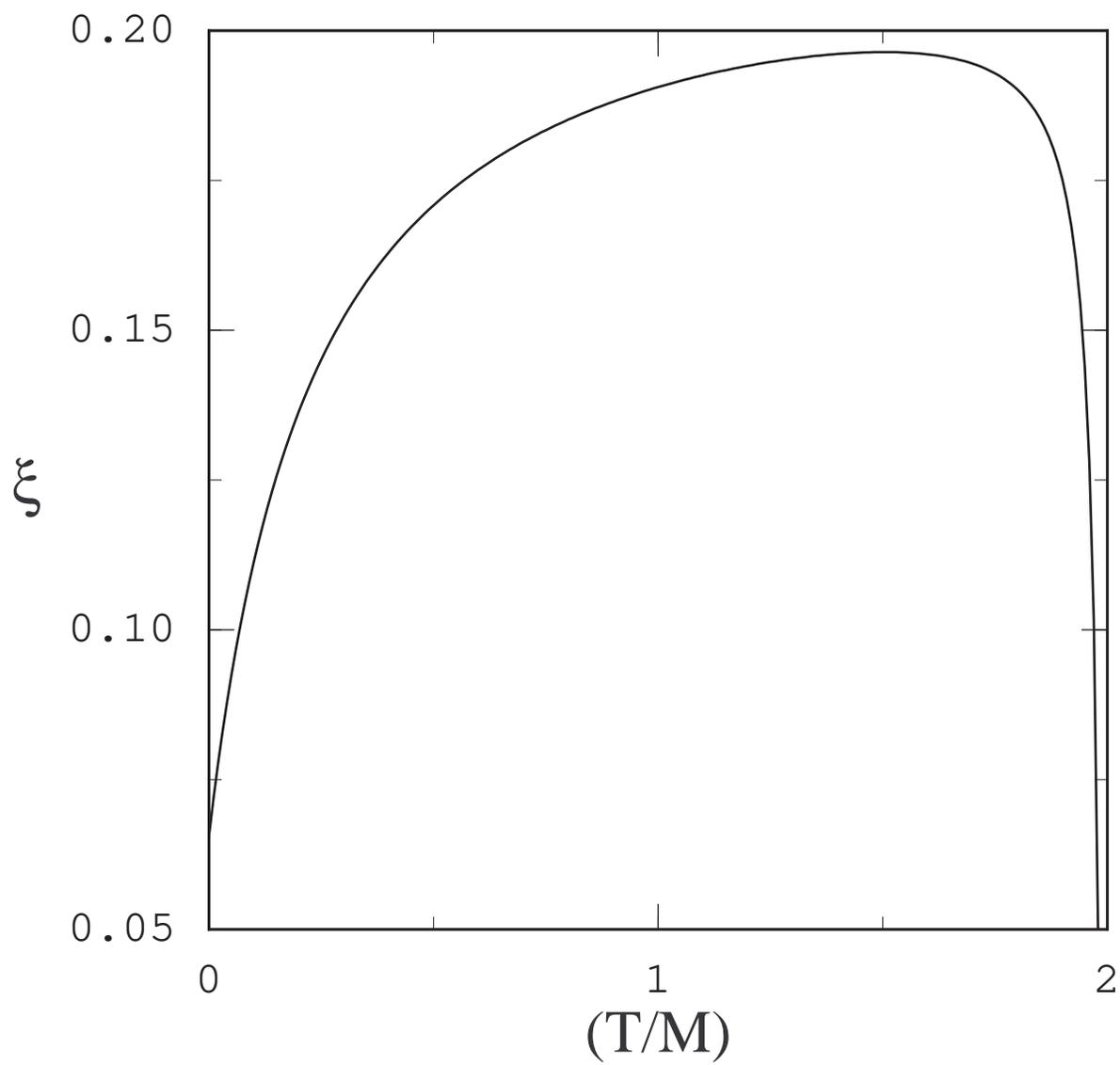
\end{document}